\documentclass[twocolumn,showpacs,amsmath,amssymb,superscriptaddress,aps,pre]{revtex4-1}
\pdfoutput=1
\usepackage{graphicx,color} 
\usepackage{dcolumn}
\usepackage{bm,bbold,empheq,mathdots}
\usepackage{amsmath}
\usepackage{graphics}
\usepackage{units,enumitem}
\usepackage{hyperref}
\usepackage{eqnarray}
\def\mydoubleq#1{``#1''}
\newcommand{\ba}{\begin{eqnarray}}
\newcommand{\ea}{\end{eqnarray}}
\newcommand{\be}{\begin{equation}}
\newcommand{\ee}{\end{equation}}
\usepackage{amsmath}    
\usepackage{graphicx}   
\usepackage{verbatim}   
\usepackage{color}      
\usepackage{subfigure}  
\usepackage{hyperref}   
\def\mydoubleq#1{``#1''}

\begin{document}

\title{Tensile elasticity of semiflexible polymers with hinge defects }
\author{Panayotis Benetatos}  \email{pben@knu.ac.kr}  \affiliation{Department of Physics, Kyungpook National University, 80 Daehakro, Bukgu, Daegu 41566, Korea}

\date{\today}  
 
\begin{abstract}

It has become clear in recent years that the simple uniform wormlike chain model needs to be modified in order to account for more complex behavior which has been observed experimentally in some important biopolymers. For example, the large flexibility of short ds-DNA has been attributed to kink or hinge defects.  In this paper, we calculate analytically, within the weak bending approximation, the force-extension relation of a wormlike chain with a permanent hinge defect along its contour. The defect is characterized by its bending energy (which can be zero, in the completely flexible case) and its position along the polymer contour. Besides the bending rigidity of the chain, these are the only parameters which describe our model. We show that a hinge defect causes a significant increase in the differential tensile compliance of a pre-stressed chain. In the small force limit, a hinge defect significantly increases the entropic elasticity.  Our results apply to any pair of semiflexible segments connected by a hinge. As such, they may also be relevant  to cytoskeletal filaments (F-actin, microtubules), where one may treat the cross-link connecting two filaments as a hinge defect.

  \end{abstract} 


\maketitle

\section{Introduction}

The wormlike chain (WLC) is a minimal theoretical model of semiflexibe polymers \cite{Kratky_Porod,STY}. It is a locally inextensible, one-dimensional, fluctuating line with bending rigidity. The latter is the single parameter of this continuous model. Despite its simplicity, it has proven quite successful in describing the entropic elasticity of long (compared with the persistence length) ds-DNA molecules \cite{Bustamante_2000}. Some important biopolymers, however, exhibit behavior which cannot be accounted for by the simple uniform and isotropic WLC model. One example is the spontaneous curvature in tubulin protofilaments, in bacterial FtsZ or in some cases of eukareotic DNA \cite{Ghosh_Singh_Sain_PRE,PB_EMT,Helmut_EPL}. In addition, starting with the pioneering experimental work of Cloutier and Widom, it has become clear that short ds-DNA molecules \cite{Cloutier_Widom} exhibit highly bendable behavior on short length scales \cite{Vafabakhsh_Ha,PNelson_rod}. Such behavior may be due to locally melted regions of the base pair sequence known as denaturation bubbles \cite{Sung,Theodorakopoulos,Menon} or to single-stranded gaps known as nicks \cite{Vologodskii}. A thermodynamically induced localization of bending has been proposed with the kinkable chain model, where kinks can, in principle, occur at any point along the chain contour \cite{Wiggins_kinkable}. For many purposes, bubbles and nicks can be viewed as hinge defects of the WLC. In \cite{Chen_Yan_PRE,Yan_Kawamura_Marko_PRE}, a transfer matrix approach is used in a discrete version of the WLC model  in order to determine the conformational and elastic effect of hinge defects. The advantage of that approach is that it is not restricted to weakly bending conformations. However, it introduces an extra parameter (the length of a link in the chain) and it does not yield closed expressions. Apart from hinge defects in ds-DNA, another motivation for the present study comes from cross-linked semiflexible polymer networks, such as reconstituted networks of cytoskeletal biopolymers \cite{Gardel}.  At least at the level of modeling, it is very common to treat cross-links as soft hinges \cite{Broedersz_RMP}. Since the dangling ends are usually ignored, two semiflexible chains end-linked by a hinge is a minimal structural element of interest for those complex systems.

In this paper, we consider a WLC with a permanent hinge defect and calculate the force-extension relation analytically in the weak bending approximation. This approximation renders the relevant functional integrals Gaussian and allows us to obtain results in closed form. Our calculations are relevant not only to ds-DNA with hinge defects, but to any hinged pair of semiflexible polymers. In fact, since the effect of a hinge is more pronounced for stiffer chains (within the weak bending approximation), our results may be more relevant to cytoskeletal filaments.

The paper is organized as follows: In Section II we calculate the force-extension relation of a WLC for two types of hinged-hinged boundary conditions: free and with constrained transverse position. This calculation introduces the formalism used in our analysis and illustrates the role of the overall tilting entropy which accounts for the elastic effect of hinge defects. In Section III, we calculate the force-extension relation for a WLC with a hinge defect of arbitrary bending stiffness (energy) located at an arbitrary fixed position along the chain's contour length.  The calculation is done for free hinged-hinged boundary conditions and is repeated in Section IV for constrained hinged-hinged boundary conditions.  In order to gain some intuitive understanding of our analytical results, we analyze the weak and strong force limits in Section V. We conclude and discuss further extensions of this work in Section VI. Details of the calculation of correlators are given in the Appendices.

\section{Two types of hinged-hinged boundary conditions}

\subsection{Free hinged-hinged boundary conditions}

We consider a semiflexible polymer modelled as a wormlike chain, stretched by a tensile force ${\bf f}$ applied at its end points. The Hamiltonian (free energy functional) reads
\begin{equation}
\label{Hamiltonian1}
{\cal H}(\{{\bf t}(s)\})=\frac{1}{2} \kappa \int_0^L\Big(\frac{d {\bf t}}{ds}\Big)^2 ds - {\bf f}\cdot\int_0^L {\bf t} ds\;,
\end{equation}
where ${\bf t}(s)$ is the tangent vector at arc-length position $s$, $L$ is the contour length, and $\kappa$ is the bending stiffness related to the persistence length $L_p$ via $L_p=\kappa /(k_B T)$ (in 3 dimensions). The local inextensibility constraint of the WLC implies that $|{\bf t}(s)|=1$. We treat the problem in the weak bending approximation, where the component of ${\bf t}(s)$  perpendicular to the direction of ${\bf f}$ is small, $t_{\perp}(s)\ll 1$. This approximation holds when the stretching force is sufficiently large or when the persistence length is much greater than the contour length. In that approximation, using the Monge parametrization of the chain, ${\bf t}(s)=(1, a_1(s), a_2(s))/\sqrt{1 + a_1^2(s)+a_2^2(s)}$, we obtain the quadratic Hamiltonian
\begin{equation}
\label{Hamiltonian2}
{\cal H}(\{a_i(s)\})=\sum_{i=1,2}\Big[\frac{1}{2} \kappa \int_0^L\big(\frac{d { a_i}}{ds}\big)^2 ds +\frac{1}{2} f \int_0^La_i^2 ds\Big] -fL\;,
\end{equation}
where we have taken the stretching force ${\bf f}$ to be in the $x$-direction. We notice that, in the weak bending approximation, the two transverse directions decouple. In the following, for the sake of simplicity, we analyze the problem with one transverse direction and we obtain the force-extension relation of the three-dimensional case by inserting a factor of 2 where needed.

In hinged-hinged boundary conditions, the bending moment at the end points vanishes\\
\begin{equation}
\label{h-h_bcs}
\frac{d a_i}{ds}\Big|_{s=0}=\frac{d a_i}{ds}\Big|_{s=L}=0\;.
\end{equation}
The appropriate Fourier decomposition of the tangent vector which satisfies the boundary conditions is a series of cosines\\
\begin{equation}
\label{Fourier1}
a(s)=\sum_{l=0}^{\infty}A_l \cos(q_l s), \;\;\;q_l=\frac{l \pi}{L}\;.
\end{equation}
Apart from a constant, the Hamiltonian of the $(1+1)$-dimensional system reads\\
\begin{equation}
\label{Hamiltonian3}
{\cal H}(\{ A_l \})=\frac{1}{2} f A_0^2 L +\sum_{l=1}^{\infty} \frac{L}{4} (\kappa q_l^2 +f) A_l^2 \;.
\end{equation}

From the equipartition theorem, we readily obtain\\
\begin{equation}
\langle A_0^2 \rangle =\frac{k_B T}{L f}
\end{equation}
and
\begin{equation}
\langle A_l^2 \rangle =\frac{2 k_B T}{L (f+ \kappa q_l^2)},\;\;\;l\neq 0
\end{equation}
The force-extension relation is obtained from\\
\begin{equation}
\label{force-ext-general}
\langle x(L) \rangle= L-\frac{1}{2}\int_0^L ds \langle t_x^2(s) +t_y^2(s)\rangle\;,
\end{equation}
where we have chosen a coordinate system such that $x(0)=0$. The correlators calculated above yield\\
\begin{equation}
\label{force-ext-h-h-free}
\frac{\langle x(L) \rangle}{L}= 1-\frac{L}{2 L_p}\frac{\coth\Big(\sqrt{\tilde{f}}\Big)\sqrt{\tilde{f}}+1}{\tilde{f}}\;,
\end{equation}
where $\tilde{f}=fL^2/\kappa$. The same result can be obtained using the method of Green functions \cite{Hori,Navid_motor}.

\subsection{Hinged-hinged boundary conditions with position constraint}

The boundary conditions discussed in the previous subsection allow for an overall end-to-end tilt of the polymer about the stretching direction. This freedom can be restricted by requiring the two end points to be at the same transverse position,
\begin{equation}
\label{pos_constr}
\int_0^L a_i(s)=0\;.
\end{equation}
Using the Fourier expansion in cosines (Eq. (\ref{Fourier1})), we get: $A_0=0$. The force-extension relation is obtained as with the free hinged-hinged boundary conditions, excluding the zeroth Fourier mode. The final result reads\\
\begin{equation}
\label{force-ext-h-h-constr}
\frac{\langle x(L) \rangle}{L}= 1-\frac{L}{2 L_p}\frac{\coth\Big(\sqrt{\tilde{f}}\Big)\sqrt{\tilde{f}}-1}{\tilde{f}}\;.
\end{equation}

The same result has been obtained using the method of Green functions \cite{Navid_motor} or by expressing the Hamiltonian in terms of the transverse displacement and Fourier expanding the latter in a series of sines in agreement with the boundary conditions \cite{PB_EMT} .

\begin{figure}
\centering
\includegraphics[
width=0.42\textwidth]{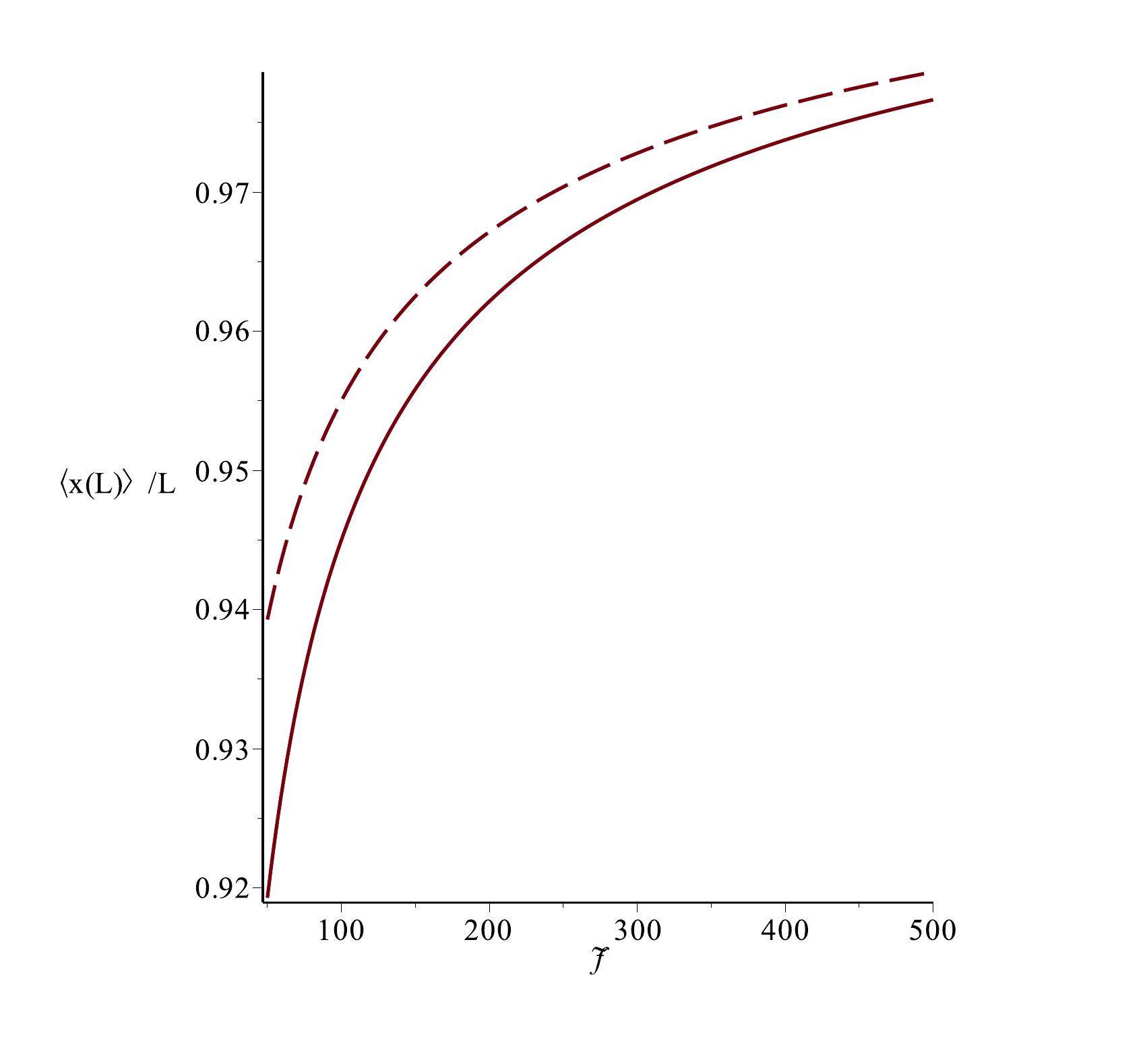}\caption{ Force-extension relation of a WLC with $L_p=L$ in two types of hinged-hinged boundary conditions: free (solid line) and with constrained transverse position (dashed line).
\label{f_ext_bcs}}
\includegraphics[
width=0.42\textwidth]{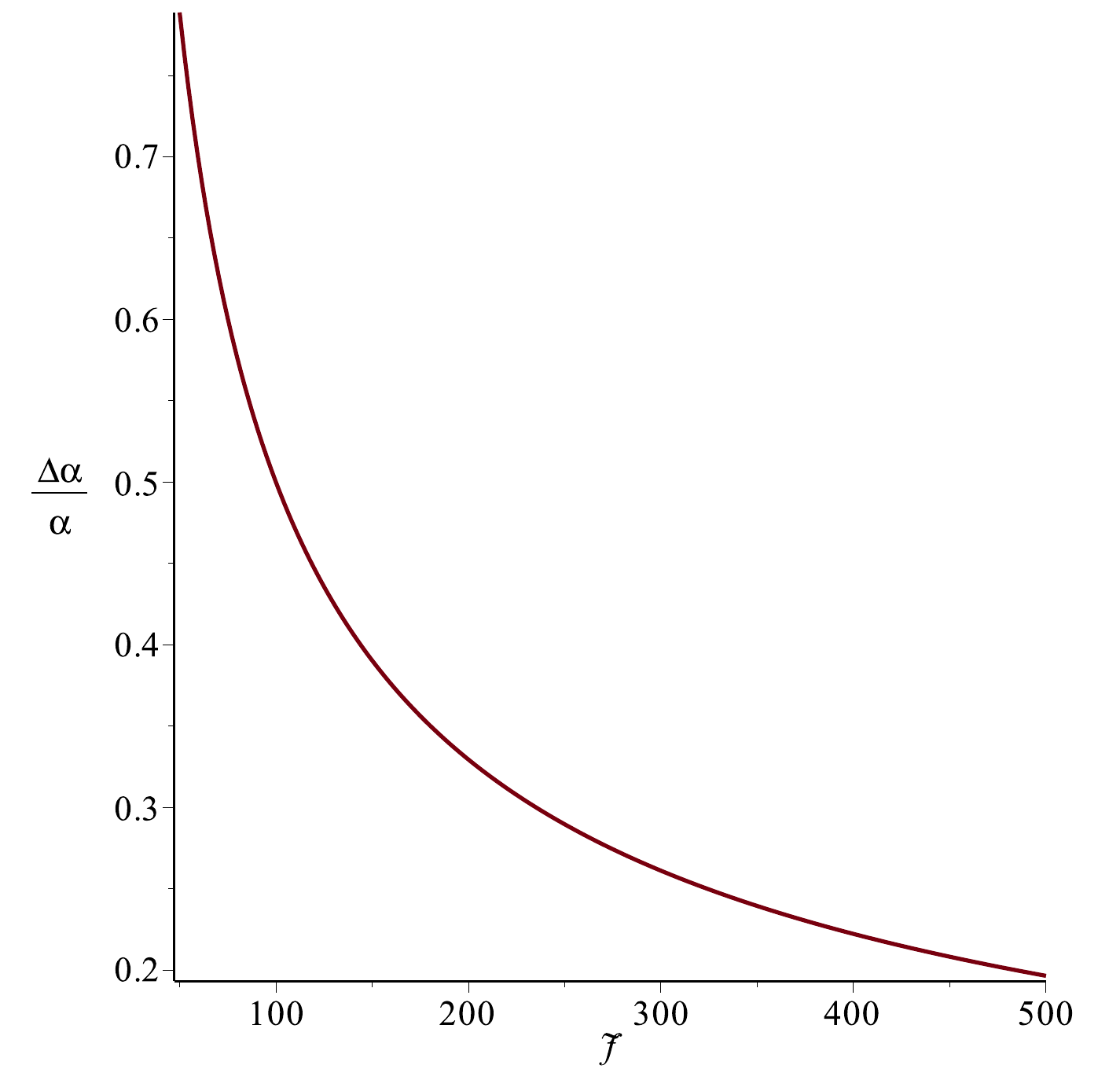}\caption{Change in the differential tensile compliance of a WLC with $L_p=L$ for free hinged-hinged boundary conditions relative to constrained hinged-hinged boundary conditions.
\label{diff_compliance_bcs}}
\end{figure}

As shown in Fig. \ref{f_ext_bcs}, for a given stretching force, the extension in the free case is smaller because the force competes with the overall tilting entropy in addition to the entropy of the thermal undulations. For $\tilde{f}\gg1$, the boundary conditions become irrelevant and we recover the Marko-Siggia result with the relative extension being inversely proportional to the square root of the force \cite{Marko-Siggia}. Even though the effect of the boundary conditions on the force-extension relation is rather small, the effect on the tensile elasticity is quite significant as shown in Fig. \ref{diff_compliance_bcs}. A measure of the tensile elasticity of the WLC is  the differential tensile compliance defined as
\begin{equation}
\label{diff_compiance}
\alpha=\frac{\partial \langle x(L)\rangle}{\partial f}\;.
\end{equation}

\section{Stretching elasticity of a WLC with a single hinge defect}

We consider a WLC with a single hinge defect modelled as a point at $s=L_1$ where the continuity of the tangent vector orientation is broken. The Hamiltonian of a stretched WLC with such a defect reads\\
\begin{align}
\label{Hamiltonian3}
& {\cal H}(\{{\bf t}_1(s)\}, \{{\bf t}_2(s)\}) = g(1-{\bf t}_1(L_1)\cdot{\bf t}_2(L_1)) \\ \nonumber
 &+ \frac{1}{2}\kappa \int_0^{L_1}\Big(\frac{d {\bf t}_1}{d s}\Big)^2 ds + \frac{1}{2}\kappa \int_{L_1}^L\big(\frac{d {\bf t}_2}{d s}\Big)^2 ds \\ \nonumber
 &+{\bf f}\cdot \int_0^{L_1} {\bf t}_1 ds + {\bf f}\cdot \int_{L_1}^{L} {\bf t}_2 ds  \;,
 \end{align}
where $g$ is an energy parameter that penalizes the misalignment of ${\bf t}_1(L_1)$ and ${\bf t}_2(L_1)$. For $g=0$ we get a completely flexible (soft) hinge. We should point out that the infinite $g$ limit does not recover the intact WLC, but yields a more flexible chain instead. In that sense, the point hinge is a genuine defect. From the way one approaches the WLC model from the discrete Kratky-Porod model \cite{Lattanzi}, we can see that in order to recover the intact WLC we need a hinge with a finite extent, $a$, so that $g\rightarrow \infty$ and $a \rightarrow 0$, keeping $ga=\kappa$ fixed. The limit $g\rightarrow \infty$, in our model, restores the continuity of the tangent vector, but it still allows for discontinuity in the curvature at $s=L_1$, rendering the WLC bending stiffness ill-defined at that point.

In the weak bending approximation, for each of the two transverse directions ($y$ and $z$) we get the following quadratic Hamiltonian
\begin{align}
\label{Hamiltonian4}
& {\cal H}(\{{a}(s)\}, \{{b}(s)\}) = \frac{1}{2}g(a(L_1)-b(L_1))^2 \\ \nonumber
 &+ \frac{1}{2}\kappa \int_0^{L_1}\Big(\frac{d {a}}{d s}\Big)^2 ds + \frac{1}{2}\kappa \int_{L_1}^L\big(\frac{d {b}}{d s}\Big)^2 ds \\ \nonumber
 &+\frac{1}{2}f\int_0^{L_1} {a}(s) ds + \frac{1}{2}f \int_{L_1}^{L} {b}(s) ds  \;.
 \end{align}
We impose hinged-hinged boundary conditions without any position constraint at $s=0,\;L_1,\;{\rm and} \;\;L\;$,
\begin{align}
\frac{d a}{ds}\Big{|}_{s=0}=\frac{d a}{ds}\Big{|}_{s={L_1}}=\frac{d b}{ds}\Big{|}_{s={L_1}}=\frac{db}{ds}\Big{|}_{s=L}=0\;.
\end{align}
This type of boundary conditions is appropriate for stretching experiments with magnetic tweezers with a freely rotating hinge at the end tethered to the substrate \cite{Chen_Yan_Biophys_J}.

We expand the fluctuating fields in the appropriate Fourier modes according to the boundary conditions\\
\begin{align}
a(s)=\sum_{l=0}^{\infty} A_l \cos(q_l s),\;\;\;q_l=\frac{l \pi}{L_1}\;,\;\;\;0< s< L_1\;,\\
b(s)=\sum_{l=0}^{\infty} B_l \cos(p_l (s-L)),\;\;p_l=\frac{l \pi}{L-L_1}\;,\;\;L_1<s<L\;.\nonumber
\end{align}
In Fourier space, the Hamiltonian functional becomes a function of the Fourier amplitudes,\\
\begin{align}
\label{Hamiltonian5}\nonumber
& {\cal H}(A_l,B_m) = \frac{1}{2}g\Big(\sum_{l,m=0}^{\infty}A_lA_m(-1)^l(-1)^m\\ \nonumber
 &+\sum_{l,m=0}^{\infty} B_lB_m(-1)^l(-1)^m -2\sum_{l,m=0}^{\infty}A_lB_m(-1)^l(-1)^m\Big)\\ \nonumber
 &+\frac{1}{2}fL_1A_0^2+ \frac{1}{2}f (L-L_1)B_0^2+\frac{1}{4}L_1\sum_{l=1}^{\infty}(\kappa q_l^2+f)A_l^2\\
 &+\frac{1}{4}(L-L_1)\sum_{l=1}^{\infty}(\kappa p_l^2+f)B_l^2\;.
 \end{align}

Introducing a column vector $\Gamma$ such that\\
\begin{align}
\Gamma^T=(A_0,B_0,A_1,B_1,...)\;,
\end{align}
a column vector $u$ such that\\
\begin{align}
u^T=\sqrt{g}(1,-1,-1,1,1,-1,-1,1,...)\;,
\end{align}
and a matrix $C$ such that\\
\begin{widetext}
\begin{gather*}
C=\frac{1}{2}
\begin{pmatrix}
2fL_1 & 0 & 0 & 0 & \hdots \\
0 & 2f(L-L_1) & 0 &  0 &\hdots \\
0 & 0 & (\kappa q_1^2+f)L_1 & 0 &  \hdots \\
0 & 0 &  0 & (\kappa p_1^2+f)(L-L_1)  \hdots \\
\vdots & \vdots & \vdots & \vdots & \ddots
\end{pmatrix}\;,
\end{gather*}
\end{widetext}
the Hamiltonian can be expressed as\\
\begin{align}
\label{Hamiltonian6}
{\cal H}(\Gamma)=\frac{1}{2}\sum_{l,m}\Gamma_l G_{lm} \Gamma_m,\;\;\;G_{lm}=C_{lm}+u_lu_m\;.
\end{align}

Since the Hamiltonian is Gaussian, correlators are calculated as follows:\\
\begin{align}\label{correlation1}
\langle \Gamma_l\Gamma_m\rangle=k_B T (G^{-1})_{lm}\;,
\end{align}
where the inverse of matrix $G$ is obtained  using the Sherman-Morrison formula from linear algebra \cite{Teukolsky},
 \begin{align}
\label{Sherman_Morrison}
 G^{-1}=C^{-1}-\frac{C^{-1}uu^TC^{-1}}{1+u^TC^{-1}u}\;.
\end{align}  
The force-extension relation is calculated from the tangent vector correlators as follows:
\begin{align}
\label{f-ext-def1}\nonumber
&\langle x(L)\rangle=L-\frac{1}{2}\int_0^{L_1}\langle a^2(s)\rangle ds - \frac{1}{2}\int_{L_1}^L\langle b^2(s)\rangle ds\nonumber\\
 & =L-\frac{L_1}{2}\langle A_0^2\rangle -\frac{L_1}{4}\sum_{l=1}^{\infty}\langle A_l^2\rangle
-\frac{L-L_1}{2}\langle B_0^2\rangle \nonumber\\
&-\frac{L-L_1}{4}\sum_{l=1}^{\infty}\langle B_l^2\rangle\;.
\end{align}
After performing some intermediate steps which are shown in Appendix A, we obtain the final result\\
\begin{widetext}
\begin{align}
\frac{\langle x(L) \rangle}{L}=&1-\frac{L}{L_p  {\tilde f}}+\frac{{\tilde g} L^4}{L_p^2 L_1(L-L_1) {\tilde f}^2\Big(1+\frac{\tilde g}{\sqrt{\tilde f}}\frac{L}{L_p}\Big(\coth\big(\sqrt{\tilde f}\frac{L_1}{L}\big)+\coth\big(\sqrt{\tilde f}\frac{(L-L_1)}{L}\big)\Big)\Big)}\nonumber\\
&-\frac{1}{2}\frac{L_1}{L_p \sqrt{\tilde f}}\coth\big(\sqrt{\tilde f}\frac{L_1}{L}\big)-\frac{1}{2}\frac{L-L_1}{L_p \sqrt{\tilde f}}\coth\big(\sqrt{\tilde f}\frac{(L-L_1)}{L}\big)\nonumber\\
&+\frac{1}{2}\frac{{\tilde g}L^3}{L_p^2 L_1 {\tilde f}^2}\frac{\frac{L_1^2}{L^2}\big(\coth\big(\sqrt{\tilde f}\frac{L_1}{L}\big)-1\big){\tilde f}+\frac{L_1}{L}\coth\big(\sqrt{\tilde f}\frac{L_1}{L}\big)\sqrt{\tilde f}-2}{1+\frac{\tilde g}{\sqrt{\tilde f}}\frac{L}{L_p}\Big(\coth\big(\sqrt{\tilde f}\frac{L_1}{L}\big)+\coth\big(\sqrt{\tilde f}\frac{(L-L_1)}{L}\big)\Big)}\nonumber\\
&+\frac{1}{2}\frac{{\tilde g}L^3}{L_p^2 (L-L_1) {\tilde f}^2}\frac{\frac{(L-L_1)^2}{L^2}\big(\coth\big(\sqrt{\tilde f}\frac{(L-L_1)}{L}\big)-1\big){\tilde f}+\frac{(L-L_1)}{L}\coth\big(\sqrt{\tilde f}\frac{(L-L_1)}{L}\big)\sqrt{\tilde f}-2}{1+\frac{\tilde g}{\sqrt{\tilde f}}\frac{L}{L_p}\Big(\coth\big(\sqrt{\tilde f}\frac{L_1}{L}\big)+\coth\big(\sqrt{\tilde f}\frac{(L-L_1)}{L}\big)\Big)}\;.
\end{align}

\end{widetext}
This final result refers to the three-dimensional case and there is a discrepancy by a factor of two with respect to the previous equation (we have two equal contributions from the two transverse directions). We can immediately see that for a soft hinge, $g=0$, the force-extension relation reduces to simply adding the contribution of two independent parts, each behaving according to Eq. (\ref{force-ext-h-h-free}). In fact, we can easily calculate the force-extension relation of a WLC with free hinged-hinged boundary conditions and an arbitrary number of soft hinges by simply adding the contributions of the individual segments. 

\begin{figure}
\centering
\includegraphics[
width=0.5\textwidth]{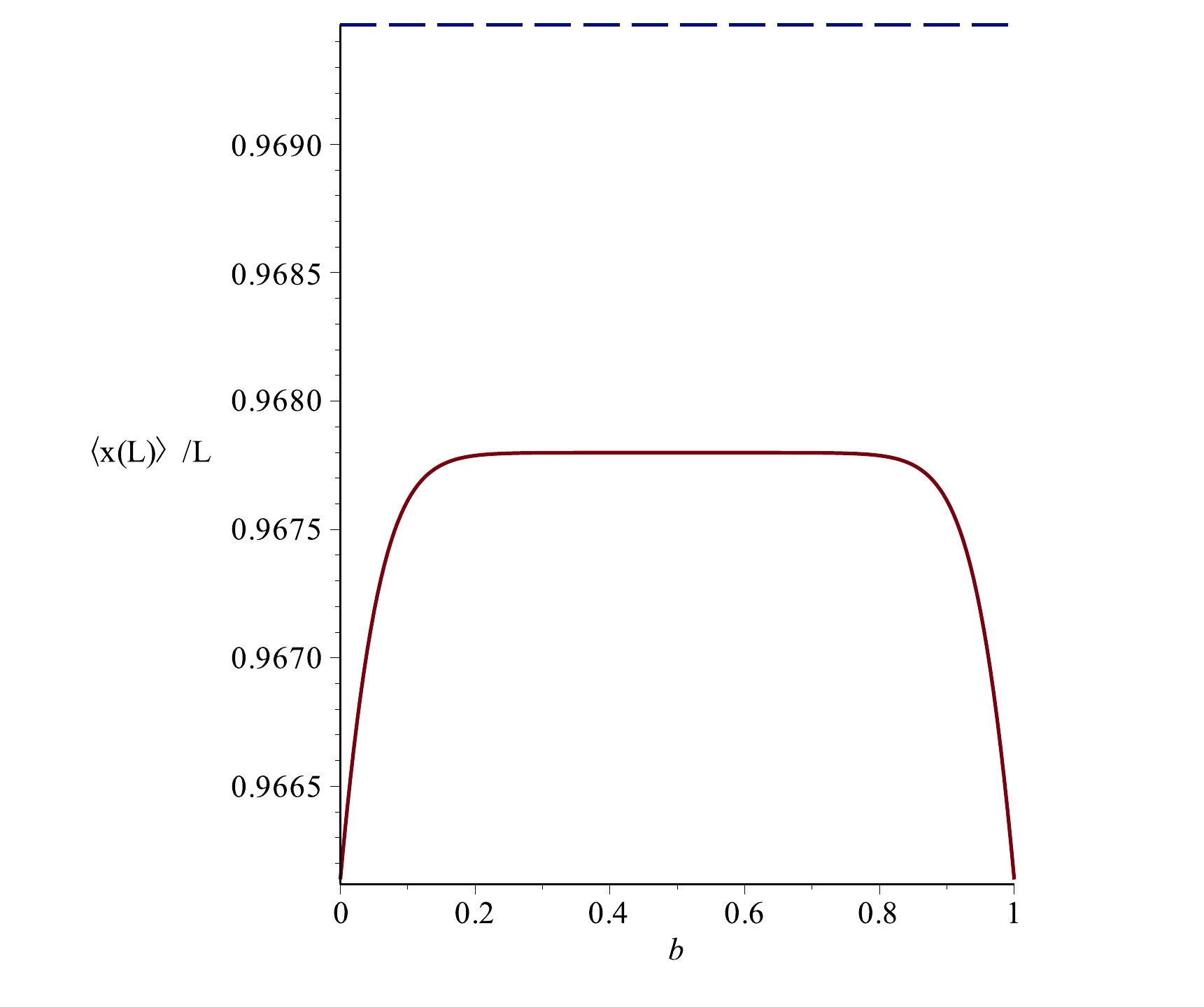}\caption{ The extension of a WLC with a soft hinge defect, stretched by a force ${\tilde f}=300$, having $L=L_p$, as a function of the defect position along the polymer contour, $b=L_1/L$. For comparison, the dashed line shows the extension of a WLC with the same parameters, without any defect. \label{defect_position}}
\end{figure}

In Fig. \ref{defect_position}, we show the dependence of the elastic response on the position of a hinge defect along the polymer contour. The response exhibits a plateau in the bulk of the chain and increases drastically as the defect enters a boundary region close to the end points whose size is given by the deflection length, $l_f=\sqrt{\kappa/f}$.
 
\begin{figure}
\centering
\includegraphics[
width=0.42\textwidth]{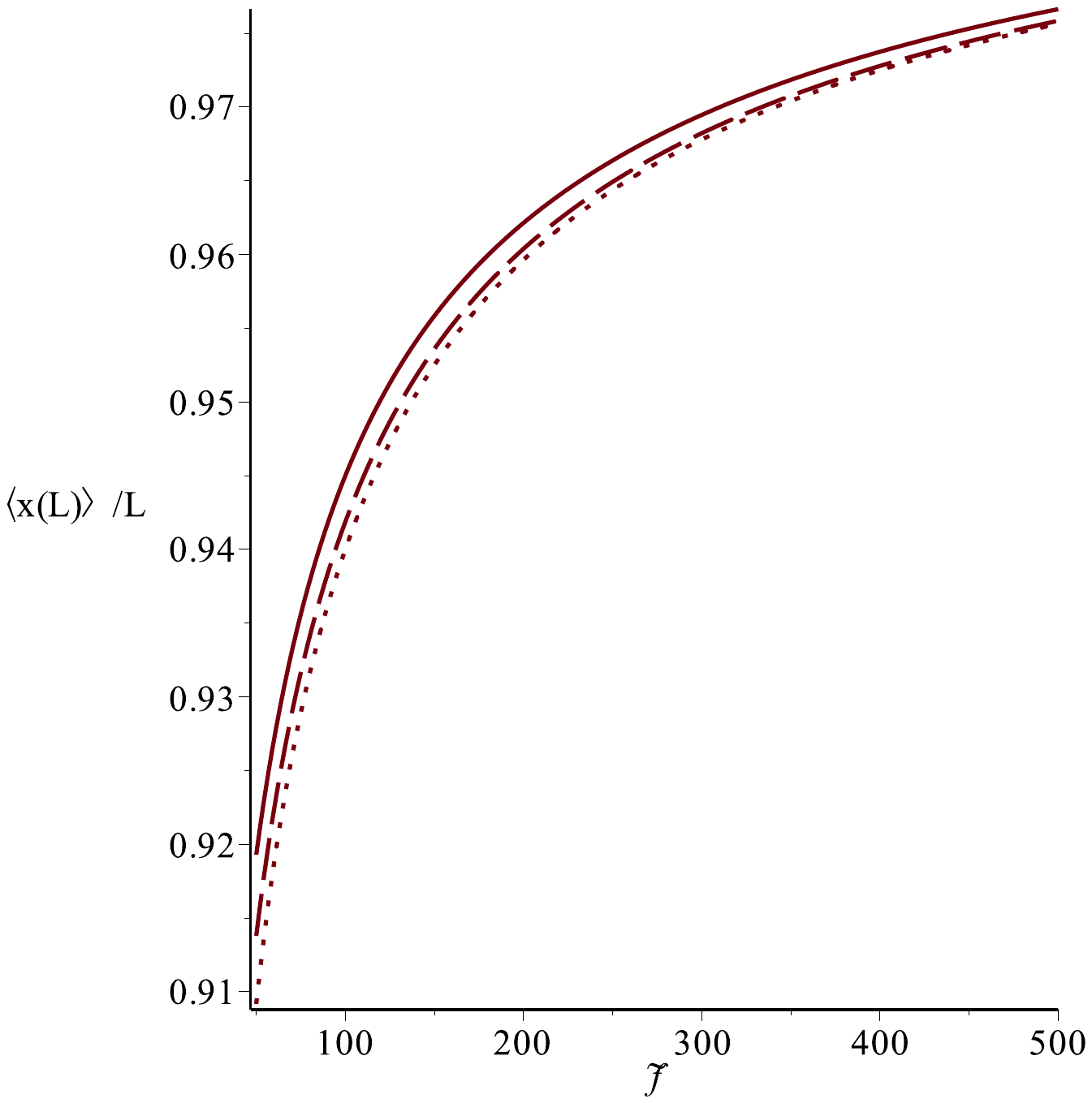}\caption{ Force-extension relation of a WLC with $L_p=L$ and free hinged-hinged boundary conditions with a soft hinge defect (dotted line), with a hinge defect with ${\tilde g}=3$ (dashed line), and without any defect (solid line). The defect is at  $L_1=L/2$. \label{f_ext_hinged}}
\includegraphics[
width=0.42\textwidth]{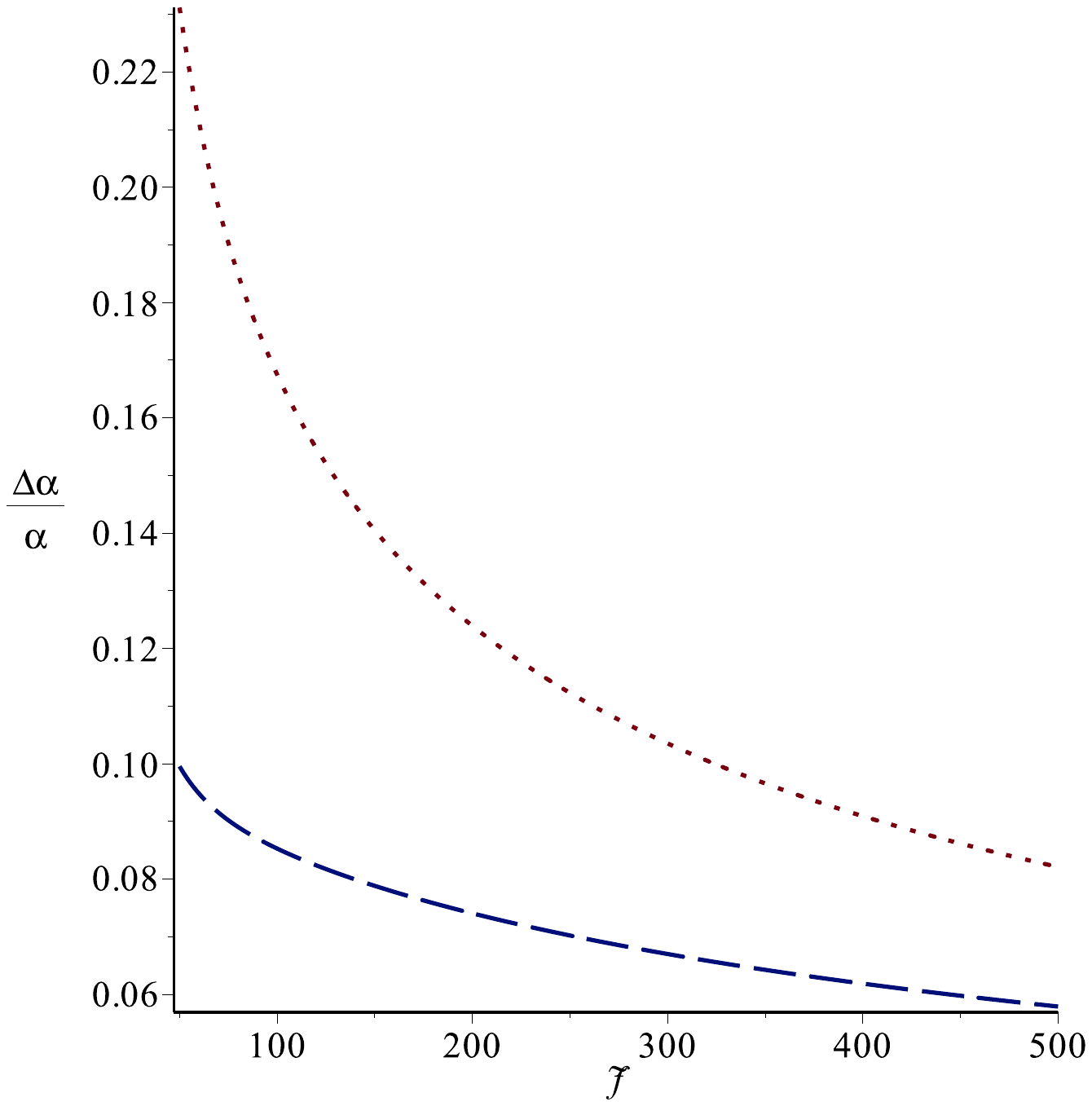}\caption{Change in the differential tensile compliance of a WLC with $L_p=L$ and  a soft hinge defect  (dotted line) or a  hinge defect with ${\tilde g}=3$ (dashed line) at the middle,  relative to that of a WLC without any defect. (Free hinged-hinged boundary conditions.)
\label{diff_compliance_hinged}}
\end{figure}

\section{The case of hinged-hinged boundary conditions with position constraint}

In this Section, we analyze the stretching elasticity of a WLC with a single hinge defect, as we did in the previous Section, but we impose a constraint in the transverse position of the hinged ends. Our motivation is  the experimental relevance  of this constraint, {\it e.g.,} in stretching experiments involving optical tweezers \cite{Purohit}. 

The requirement that the two end-points are at the same transverse position is expressed by the following equation (in $1+1$ dimensions):\\
\begin{align}
\int_0^{L_1}ds\, a(s) +\int_{L_1}^{L}ds\, b(s)=0\;,
\end{align}
which, in Fourier space, implies $A_0=-\frac{L-L_1}{L_1}B_0$. In Fourier space, the Hamiltonian reads as in Eq. (\ref{Hamiltonian6}), but now \\
\begin{align}
\Gamma^T=(A_0,A_1,B_1,A_2,B_2,....)\;,
\end{align}
\begin{align}
u^T=\sqrt{g}(\frac{L}{L-L_1},-1,1,1,-1,-1,1,...)\;,
\end{align}
and
\begin{widetext}
\begin{gather*}
C=\frac{1}{2}
\begin{pmatrix}
2fLL_1/(L-L_1)& 0 & 0 & 0 & \hdots \\
0 &(\kappa q_1^2+f)L_1  & 0 &  0 &\hdots \\
0 & 0 & (\kappa p_1^2+f)(L-L_1) & 0 &  \hdots \\
0 & 0 &  0 & (\kappa q_2^2+f)L_1  \hdots \\
\vdots & \vdots & \vdots & \vdots & \ddots
\end{pmatrix}\;.
\end{gather*}
\end{widetext}                                                                                                                                                                   

The force-extension relation is obtained from the tangent vector correlators as in Eq. (\ref{f-ext-def1}) using some intermediate steps which are shown in Appendix B. The final result reads (again, we have inserted a factor of 2 to account for the 2 transverse directions)\\
\begin{widetext}
\begin{align}
\frac{\langle x(L) \rangle}{L}=&1+\frac{{\tilde g}L^4}{(L-L_1) L_1 L_p^2 {\tilde f }^2\Big(1+\frac{\tilde g}{\sqrt{\tilde f}}\frac{L}{L_p}\Big(\coth\big(\sqrt{\tilde f}\frac{L_1}{L}\big)+\coth\big(\sqrt{\tilde f}\frac{(L-L_1)}{L}\big)\Big)\Big)}\nonumber\\
&-\frac{1}{2}\frac{L}{L_p  \sqrt{\tilde f}}\Big(\frac{L_1}{L}\coth\big(\sqrt{\tilde f}\frac{L_1}{L}\big)+\frac{L-L_1}{L}\coth\big(\sqrt{\tilde f}\frac{(L-L_1)}{L}\big)\Big)\nonumber\\
&+\frac{1}{2}\frac{{\tilde g}L^3}{L_p^2 L_1 {\tilde f}^2}\frac{\frac{L_1^2}{L^2}\big(\coth\big(\sqrt{\tilde f}\frac{L_1}{L}\big)-1\big){\tilde f}+\frac{L_1}{L}\coth\big(\sqrt{\tilde f}\frac{L_1}{L}\big)\sqrt{\tilde f}-2}{1+\frac{\tilde g}{\sqrt{\tilde f}}\frac{L}{L_p}\Big(\coth\big(\sqrt{\tilde f}\frac{L_1}{L}\big)+\coth\big(\sqrt{\tilde f}\frac{(L-L_1)}{L}\big)\Big)\;}\nonumber\\
&+\frac{1}{2}\frac{{\tilde g}L^3}{L_p^2 (L-L_1) {\tilde f}^2}\frac{\frac{(L-L_1)^2}{L^2}\big(\coth\big(\sqrt{\tilde f}\frac{(L-L_1)}{L}\big)-1\big){\tilde f}+\frac{(L-L_1)}{L}\coth\big(\sqrt{\tilde f}\frac{(L-L_1)}{L}\big)\sqrt{\tilde f}-2}{1+\frac{\tilde g}{\sqrt{\tilde f}}\frac{L}{L_p}\Big(\coth\big(\sqrt{\tilde f}\frac{L_1}{L}\big)+\coth\big(\sqrt{\tilde f}\frac{(L-L_1)}{L}\big)\Big)}
\end{align}

\end{widetext}

\begin{figure}
\centering
\includegraphics[
width=0.42\textwidth]{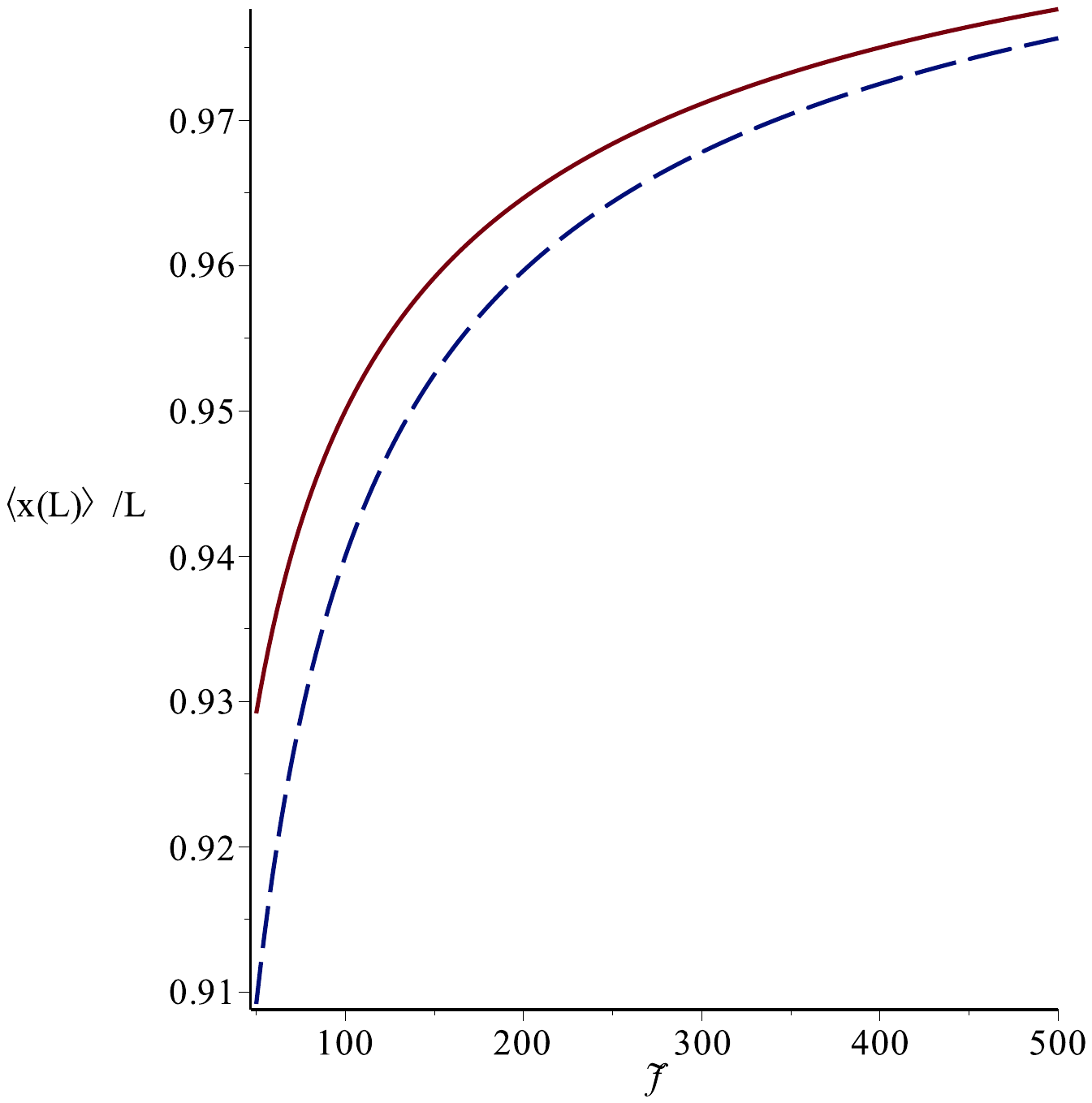}\caption{ Force-extension relation of a WLC with $L_p=L$ and  a soft hinge defect at the middle, in two types of hinged-hinged boundary conditions: free (dashed line) and with constrained transverse position (solid line).
\label{f_ext_hinged_bcs}}
\includegraphics[
width=0.42\textwidth]{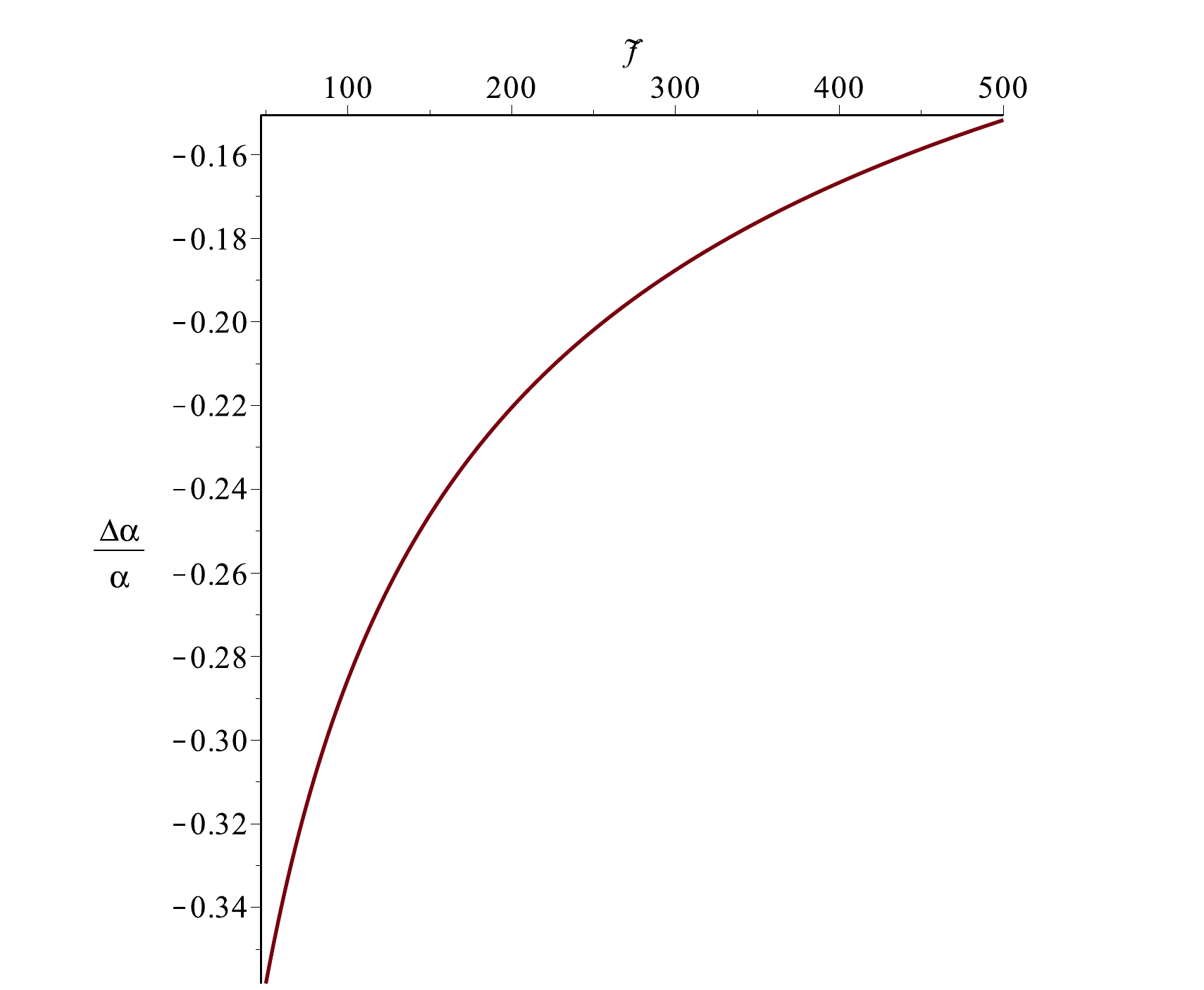}\caption{Change in the differential tensile compliance of a WLC with $L_p=L$ and  a soft hinge defect at the middle, for constrained hinged-hinged boundary conditions relative to free hinged-hinged boundary conditions.
\label{diff_compliance_hinged_bcs}}
\end{figure}

As expected, the position constraint straightens and stiffens the polymer. The elastic response is shown in Fig. \ref{f_ext_hinged_bcs} and Fig. \ref{diff_compliance_hinged_bcs}.

\section{Small and Large Force Limits}

In order to gain some intuition into the effect of a hinge on the elasticity of a WLC, in this Section, we consider the limit cases of strong and weak pulling forces. At first, we consider free hinged-hinged boundary conditions. The large force limit yields\\
\begin{equation}
\frac{\langle x \rangle}{L}=1-\frac{L}{2L_p \sqrt{\tilde f}}-\frac{L}{L_p {\tilde f}}+{\cal O}({\tilde f}^{-3/2})\;,
\end{equation}
independently of the strength of the hinge. If we compare it with the large force limit of the intact chain (without any hinge),
\begin{equation}
\frac{\langle x \rangle}{L}=1-\frac{L}{2L_p \sqrt{\tilde f}}-\frac{L}{2L_p {\tilde f}}\;,
\end{equation}
we see that hinge defect changes the subdominant term (inversely proportional to the force) by a factor of two. This term, which is purely entropic, expresses the elasticity of a freely jointed chain at the strong stretching limit \cite{Rubinstein}. At the opposite limit of a small pulling force, we obtain\\
\begin{equation}
\frac{\langle x \rangle}{L}=1-\frac{3}{2}\frac{L}{L_p {\tilde f}}=1-\frac{3}{2}\frac{k_B T}{L f}\;,
\end{equation}
independently of the strength of the hinge. If we compare it with the small force limit of the intact chain,
\begin{equation}
\frac{\langle x \rangle}{L}=1-\frac{L}{L_p {\tilde f}}=1-\frac{k_B T}{L f}\;,
\end{equation}
we see that the hinge increases the entropic elasticity by a factor of $3/2$. Of course, for the small force limit to be consistent with the weak bending approximation, we need polymers with large persistence length. This limit may be relevant for cytoskeletal fibers (F-actin, microtubules).

We now examine the case of constrained hinged-hinged boundary conditions. The large force limit, $f\gg \max(\frac{\kappa}{L^2}, \frac{g^2}{\kappa})$, yields\\
\begin{equation}
\frac{\langle x \rangle}{L}=1-\frac{L}{2L_p \sqrt{\tilde f}}+\frac{{\tilde g} L^2}{L_p^2 {\tilde f}^{3/2}}+{\cal O}({\tilde f}^{-2})\;,
\end{equation}
which should be compared with the corresponding result from the intact chain\\
\begin{equation}
\frac{\langle x \rangle}{L}=1-\frac{L}{2L_p \sqrt{\tilde f}}+\frac{L}{2L_p {\tilde f}}\;.
\end{equation}
We see that a hinge with finite stiffness replaces  the subdominant (inversely proportional to the force)  term in the force-extension relation by another subdominant term (inversely proportional to $f^{3/2}$ and proportional to the hinge stiffness).

For constrained hinged-hinged boundary conditions, the small force limit of the intact chain  is interesting:\\
\begin{equation}
\frac{\langle x \rangle}{L}=1-\frac{1}{6}\frac{L}{L_p}+\frac{1}{90}\frac{L}{L_p}{\tilde f}\;.
\end{equation}
There is a linear response to both tensile ($f>0$) and compressional ($f<0$) forces. The existence of a compressional linear response indicates the cancellation of the Euler buckling instability due to the thermal undulations (\mydoubleq{thermodynamic buckling}, \cite{Klaus_Euler}). It is remarkable that the presence of a hinge defect, irrespective of its strength, cancels the linear response and yields a nonlinear response instead,\\
\begin{equation}
\frac{\langle x \rangle}{L}=1-\frac{L}{2 L_p {\tilde f}}=1-\frac{k_B T}{2 L f}\;.
\end{equation}
If we compare this equation with the corresponding equation for free hinged-hinged boundary conditions, we see that the position constraint causes a reduction in the entropic elasticity by a factor of three.

                                   \section{Discussion - Conclusions}
             
             In this paper, we investigated the effect of a permanent hinge defect on the tensile elasticity of a WLC. We treated the problem in the weak bending approximation, and we obtained the force-extension relation for two types of boundary conditions corresponding to different types of experiments. For given  total contour length and temperature, our result depends on the hinge position, its bending energy, and the bending rigidity of the original WLC. We find that a hinge defect shortens the end-to-end distance of the stretched WLC and, at the same time, increases its differential tensile compliance. It is remarkable that this behavior does not change, irrespective of whether the hinge is completely flexible or it has a large bending energy. For the defect to cause a significant shift in the force-extension relation, the deflection length, $l_f=\sqrt{{\kappa}/f}$, should be of the order of the total contour length or greater. This is expected, as for $l_f\ll L$, the WLC can be viewed as an effective freely jointed chain consisting of links, each of length $l_f$ \cite{PB_EMT}. We point out that the shift in the differential compliance caused by the defect is much more pronounced than the shift in the force-extension relation. In the strong stretching regime, the position of the hinge does not affect the elastic response, except for a boundary region at the end points, of the size of the deflection length. In the small force limit, which is amenable to our analysis for rather stiff chains, the effect of the hinge on the elasticity is always significant. Remarkably, it destroys the linear response which is known to exist in the case of position-constrained boundary conditions. Our results may prove useful in the interpretation of stretching experiments of semiflexible biopolymers with hinge defects. Because of the above-mentioned deflection-length condition, our results may be more relevant to the study of hinged cytoskeletal filaments.
  
  From the methodological point of view, our work illustrates the usefulness of the Sherman-Morrison formula in extending a Gaussian theory to incorporate non-trivial features. In \cite{PB_Ulrich_NJP} and \cite{PB_shrinkage}, it was used in order to describe a polymer crosslink.
             
                 Even though permanent ds-DNA bubbles are possible \cite{vonHippel,Corbett}, \mydoubleq{breathing} bubbles are transient \cite{breathingDNA}. An interesting extension of our work could take into account such  transient hinge defects. Other possible directions for future work would be the analysis of hinge defects of finite extent, and the interplay of hinge defects with the twist elasticity of a polymer under tension.            

\begin{widetext}

                                                                                                                        \appendix
     
     \section{Correlators and sums used in the force-extension relation (free hinged-hinged boundary conditions)}                                                                                                                                                                 

The denominator in the Sherman-Morrison formula, Eq. \ref{Sherman_Morrison} is\\
\begin{align}
1+u^TC^{-1}u&=1+2g\Big(\frac{1}{L_1}\sum_{l=1}^{\infty}\frac{1}{\kappa q_l^2+f}
+\frac{1}{2L_1 f}+\frac{1}{L-L_1}\sum_{l=1}^{\infty}\frac{1}{\kappa p_l^2+f}+\frac{1}{2(L-L_1)f}\Big)\nonumber\\
&=1+\frac{\tilde g}{\sqrt{\tilde f}}\frac{L}{L_p}\Big(\coth\big(\sqrt{\tilde f}\frac{L_1}{L}\big)+\coth\big(\sqrt{\tilde f}\frac{(L-L_1)}{L}\big)\Big)\;,
\end{align}
where ${\tilde g}=g/(k_B T)$.

From the Sherman-Morrison formula, we obtain the correlators:\\
\begin{align}
\langle A_0^2 \rangle =\frac{L^2}{L_p L_1 {\tilde f}}-\frac{{\tilde g} L^4}{L_p^2 L_1^2 {\tilde f}^2\Big(1+\frac{\tilde g}{\sqrt{\tilde f}}\frac{L}{L_p}\Big(\coth\big(\sqrt{\tilde f}\frac{L_1}{L}\big)+\coth\big(\sqrt{\tilde f}\frac{(L-L_1)}{L}\big)\Big)\Big)}\;,
\end{align}
 \begin{align}
\sum_{l=1}^{\infty}\langle A_l^2 \rangle &= k_B T\Big(\frac{2}{L_1}\sum_{l=1}^{\infty}\frac{1}{\kappa q_l^2 +f }-\frac{g}{1+\frac{\tilde g}{\sqrt{\tilde f}}\frac{L}{L_p}\Big(\coth\big(\sqrt{\tilde f}\frac{L_1}{L}\big)+\coth\big(\sqrt{\tilde f}\frac{(L-L_1)}{L}\big)\Big)}\frac{4}{L_1^2}\sum_{l=1}^{\infty}\frac{1}{(\kappa q_l^2 +f)^2}\Big)\nonumber\\
&=\frac{L^2}{L_p L_1 {\tilde f}}\Big(\frac{L_1}{L}\coth\big(\sqrt{\tilde f}\frac{L_1}{L}\big)\sqrt{\tilde f}-1\Big)-\frac{{\tilde g}L^4}{L_p^2 L_1^2 {\tilde f}^2}\frac{\frac{L_1^2}{L^2}\big(\coth\big(\sqrt{\tilde f}\frac{L_1}{L}\big)-1\big){\tilde f}+\frac{L_1}{L}\coth\big(\sqrt{\tilde f}\frac{L_1}{L}\big)\sqrt{\tilde f}-2}{1+\frac{\tilde g}{\sqrt{\tilde f}}\frac{L}{L_p}\Big(\coth\big(\sqrt{\tilde f}\frac{L_1}{L}\big)+\coth\big(\sqrt{\tilde f}\frac{(L-L_1)}{L}\big)\Big)}\;,
 \end{align}
 \begin{align}
 \langle B_0^2 \rangle =\frac{L^2}{L_p (L-L_1) {\tilde f}}-\frac{{\tilde g} L^4}{L_p^2 (L-L_1)^2 {\tilde f}^2\Big(1+\frac{\tilde g}{\sqrt{\tilde f}}\frac{L}{L_p}\Big(\coth\big(\sqrt{\tilde f}\frac{L_1}{L}\big)+\coth\big(\sqrt{\tilde f}\frac{(L-L_1)}{L}\big)\Big)\Big)}\;,
\end{align}
\begin{align}
\sum_{l=1}^{\infty}\langle B_l^2 \rangle &= k_B T\Big(\frac{2}{L-L_1}\sum_{l=1}^{\infty}\frac{1}{\kappa p_l^2 +f }-\frac{g}{1+\frac{\tilde g}{\sqrt{\tilde f}}\frac{L}{L_p}\Big(\coth\big(\sqrt{\tilde f}\frac{L_1}{L}\big)+\coth\big(\sqrt{\tilde f}\frac{(L-L_1)}{L}\big)\Big)}\frac{4}{(L-L_1)^2}\sum_{l=1}^{\infty}\frac{1}{(\kappa p_l^2 +f)^2}\Big)\nonumber\\
&=\frac{L^2}{L_p (L-L_1) {\tilde f}}\Big(\frac{L-L_1}{L}\coth\big(\sqrt{\tilde f}\frac{(L-L_1)}{L}\big)\sqrt{\tilde f}-1\Big)\nonumber\\
&-\frac{{\tilde g}L^4}{L_p^2 (L-L_1)^2 {\tilde f}^2}\frac{\frac{(L-L_1)^2}{L^2}\big(\coth\big(\sqrt{\tilde f}\frac{(L-L_1)}{L}\big)-1\big){\tilde f}+\frac{(L-L_1)}{L}\coth\big(\sqrt{\tilde f}\frac{(L-L_1)}{L}\big)\sqrt{\tilde f}-2}{1+\frac{\tilde g}{\sqrt{\tilde f}}\frac{L}{L_p}\Big(\coth\big(\sqrt{\tilde f}\frac{L_1}{L}\big)+\coth\big(\sqrt{\tilde f}\frac{(L-L_1)}{L}\big)\Big)}\;.
 \end{align}

  \section{Correlators and sums used in the force-extension relation (constrained hinged-hinged boundary conditions)}                                                                                                                                                                 

The denominator of the Sherman-Morrison formula reads:\\
 \begin{align}
1+u^TC^{-1}u&=1+g\Big(\frac{1}{L_1}\sum_{l=1}^{\infty}\frac{2}{\kappa q_l^2+f}
+\frac{2}{L-L_1}\sum_{l=1}^{\infty}\frac{1}{\kappa p_l^2+f}+\frac{L}{(L-L_1)L_1 f}\Big)\nonumber\\
&=1+\frac{\tilde g}{{\tilde f}}\frac{L}{L_p}\Big(\coth\big(\sqrt{\tilde f}\frac{L_1}{L}\big)\sqrt{\tilde f}+\coth\big(\sqrt{\tilde f}\frac{(L-L_1)}{L}\big)\sqrt{\tilde f}\Big)\;.
\end{align}
We also get:\\
 \begin{align}
\langle A_0^2 \rangle =\Big(\frac{L-L_1}{L_1}\Big)^2 \langle B_0^2 \rangle =\frac{L (L-L_1)}{L_1L_p {\tilde f}}- \frac{{\tilde g}L^4 }{L_1^2 L_p^2 {\tilde f }^2\Big(1+\frac{\tilde g}{{\tilde f}}\frac{L}{L_p}\Big(\coth\big(\sqrt{\tilde f}\frac{L_1}{L}\big)\sqrt{\tilde f}+\coth\big(\sqrt{\tilde f}\frac{(L-L_1)}{L}\big)\sqrt{\tilde f}\Big)\Big)}\;,
  \end{align}
 \begin{align}
 &\sum_{l=1}^{\infty}\langle A_l^2 \rangle  = k_B T\Big(\frac{2}{L_1}\sum_{l=1}^{\infty}\frac{1}{\kappa q_l^2 +f }-\frac{g}{1+\frac{\tilde g}{{\tilde f}}\frac{L}{L_p}\Big(\coth\big(\sqrt{\tilde f}\frac{L_1}{L}\big)\sqrt{\tilde f}+\coth\big(\sqrt{\tilde f}\frac{(L-L_1)}{L}\big)\sqrt{\tilde f}\Big)}\frac{4}{L_1^2}\sum_{l=1}^{\infty}\frac{1}{(\kappa q_l^2 +f)^2}\Big)\nonumber\\
&=\frac{L^2}{L_p L_1 {\tilde f}}\Big(\frac{L_1}{L}\coth\big(\sqrt{\tilde f}\frac{L_1}{L}\big)\sqrt{\tilde f}-1\Big)-\frac{{\tilde g}L^4}{L_p^2 L_1^2 {\tilde f}^2}\frac{\frac{L_1^2}{L^2}\big(\coth\big(\sqrt{\tilde f}\frac{L_1}{L}\big)-1\big){\tilde f}+\frac{L_1}{L}\coth\big(\sqrt{\tilde f}\frac{L_1}{L}\big)\sqrt{\tilde f}-2}{1+\frac{\tilde g}{{\tilde f}}\frac{L}{L_p}\Big(\coth\big(\sqrt{\tilde f}\frac{L_1}{L}\big)\sqrt{\tilde f}+\coth\big(\sqrt{\tilde f}\frac{(L-L_1)}{L}\big)\sqrt{\tilde f}\Big)\;}\;,
 \end{align} 
 \begin{align}
\sum_{l=1}^{\infty}\langle B_l^2 \rangle &= k_B T\Big(\frac{2}{L-L_1}\sum_{l=1}^{\infty}\frac{1}{\kappa p_l^2 +f }\nonumber\\
&-\frac{g}{1+\frac{\tilde g}{{\tilde f}}\frac{L}{L_p}\Big(\coth\big(\sqrt{\tilde f}\frac{L_1}{L}\big)\sqrt{\tilde f}+\coth\big(\sqrt{\tilde f}\frac{(L-L_1)}{L}\big)\sqrt{\tilde f}\Big)}\frac{4}{(L-L_1)^2}\sum_{l=1}^{\infty}\frac{1}{(\kappa p_l^2 +f)^2}\Big)\nonumber\\
&=\frac{L^2}{L_p (L-L_1) {\tilde f}}\Big(\frac{L-L_1}{L}\coth\big(\sqrt{\tilde f}\frac{(L-L_1)}{L}\big)\sqrt{\tilde f}-1\Big)\nonumber\\
&-\frac{{\tilde g}L^4}{L_p^2 (L-L_1)^2 {\tilde f}^2}\frac{\frac{(L-L_1)^2}{L^2}\big(\coth\big(\sqrt{\tilde f}\frac{(L-L_1)}{L}\big)-1\big){\tilde f}+\frac{(L-L_1)}{L}\coth\big(\sqrt{\tilde f}\frac{(L-L_1)}{L}\big)\sqrt{\tilde f}-2}{1+\frac{\tilde g}{{\tilde f}}\frac{L}{L_p}\Big(\coth\big(\sqrt{\tilde f}\frac{L_1}{L}\big)\sqrt{\tilde f}+\coth\big(\sqrt{\tilde f}\frac{(L-L_1)}{L}\big)\sqrt{\tilde f}\Big)}\;.
 \end{align}

  \end{widetext}    
                                                                                                                                                                                                                                                                       
\begin{acknowledgements}

I thank Dr. Nikos Theodorakopoulos for motivating discussions, and the the Department of Solid State Physics at the National and Kapodistrian University of Athens for hospitality during part of this work.
\end{acknowledgements}

\end{document}